\documentclass[conference]{IEEEtran}
\IEEEoverridecommandlockouts
\usepackage{cite}
\usepackage{amsmath,amssymb,amsfonts}
\usepackage{algorithmic}
\usepackage{graphicx}
\usepackage{textcomp}
\usepackage{xcolor}
\usepackage{algorithm}
\usepackage{color,soul}
\usepackage{hyperref}

\usepackage[shortlabels]{enumitem}
\def\BibTeX{{\rm B\kern-.05em{\sc i\kern-.025em b}\kern-.08em
    T\kern-.1667em\lower.7ex\hbox{E}\kern-.125emX}}

\begin{document}

\title{Provider-centric Allocation of Drone Swarm Services \vspace{-0.5cm}}

\author{\IEEEauthorblockN{1\textsuperscript{st} Balsam Alkouz}
\IEEEauthorblockA{\textit{School of Computer Science} \\
\textit{The University of Sydney}\\
Sydney, Australia \\
balsam.alkouz@sydney.edu.au}
\and
\IEEEauthorblockN{2\textsuperscript{nd} Athman Bouguettaya}
\IEEEauthorblockA{\textit{School of Computer Science} \\
\textit{The University of Sydney}\\
Sydney, Australia \\
athman.bouguettaya@sydney.edu.au}
}

\maketitle

\begin{abstract}

We propose a novel framework for the allocation of drone swarms for delivery services known as Swarm-based Drone-as-a-Service (SDaaS). The allocation framework ensures minimum cost (aka maximum profit) to drone swarm providers while meeting the time requirement of service consumers. The constraints in the delivery environment (e.g., limited recharging pads) are taken into consideration. We propose three algorithms to select the best allocation of drone swarms given a set of requests from multiple consumers. We conduct a set of experiments to evaluate and compare the efficiency of these algorithms considering the provider's profit, feasibility, requests fulfilment, and drones utilization level.
\end{abstract}


\section{Introduction}
Drone-as-a-Service (DaaS) is a concept used to describe the use of the service paradigm to model services offered by unmanned aerial vehicles \cite{shahzaad2021resilient}. These services include, but are not limited to, agriculture \cite{mogili2018review}, geographic mapping \cite{madawalagama2016low}, and package delivery \cite{shahzaad2020game}. Of particular interest, is the application of DaaS in delivery. Drone delivery has recently seen a boost in interest from the industry \cite{forum_2020} and research \cite{euchi2020drones} as the need for contact-less deliveries increases in times of pandemics. Governments are relaxing delivery restrictions and regulations which accelerates the drone delivery growth \cite{forum_2020}. Swam-based DaaS (SDaaS) augments the concept to describe drone services that are delivered by a swarm of drones instead of single drones \cite{alkouz2020swarm}. SDaaS services are widely used in search and rescue \cite{cardona2019robot}, sky shows entertainment \cite{waibel2017drone}, airborne communication networks \cite{cao2018airborne}, and delivery \cite{alkouz2020formation}. \looseness=-1

Swarms of drones in delivery are able to overcome a single drone delivery limitations. A swarm may be used for the timely delivery of heavier and/or multiple packages which go beyond the capability of one single drone \cite{alkouz2020swarm}. A swarm may also travel longer distances as the payload may be distributed amongst the drones which in return reduces the pressure on the battery \cite{alkouz2020formation}. In addition, recent USA’s Federal Aviation Administration (FAA) drone flying regulations approved the use of small drones for delivery (payload $<$ 2.5 kg)\footnote{https://www.faa.gov/uas/advanced\_operations/package\_delivery\_drone}. Therefore, larger drones are not regulated as they are neither practical nor safe in a city which raises the need for drone swarms in delivery. Swarm-based Drone-as-a-Service for \emph{delivery} is defined as the use of swarm of drones to carry multiple packages in a line of sight \emph{skyway segment} \cite{alkouz2020swarm}. The skyway segments connect nodes that maybe the source, destination, or intermediate nodes representing buildings rooftops equipped with recharging stations. The nodes and segments would make an SDaaS skyway network. SDaaS delivery composes of three main components. First, \emph{delivery swarms must be allocated for delivery requests}. Second, the best path must be composed from the source to the destination that optimizes the delivery. Last, in case of failure, recovery plans must be implemented. In this paper, we focus on the first component (allocation) taking into consideration the effect on the second component (composition). Previous works focused on the composition assuming the existence of pre-allocated swarms \cite{alkouz2020swarm}. \looseness=-1

Swarm-based drone delivery services lend themselves quite naturally to being modelled using the service paradigm because they map to the key ingredients of the service concept, i.e., functional and non-functional attributes \cite{alkouz2020swarm}. The function of an SDaaS is the delivery of packages from a source to a destination. The non-functional aspects or the Quality of Services (QoS) include the delivery time, energy consumed, cost, etc. We address the allocation of SDaaS services from a \emph{provider perspective}, i.e. optimizing the QoS for the benefit of the provider. {We focus on optimizing the profit through the allocation of swarm members to consumer's requests}. The allocation of the swarm members is a core part of swarm-based services as it depends on the consumers requests and directly affects the optimal composition of SDaaS services.

There are several challenges that require addressing when allocating swarms for delivery requests. We assume that an SDaaS service provider owns a fleet of $n$ drones that are used to make multiple swarms to serve multiple requests. The swarms need to be allocated optimally to deliver requests bounded by \emph{strict time windows}. A late or early delivery is considered unsuccessful. In addition, swarms may travel longer than a request's time window for certain delivery missions (e.g. long distances). This causes \emph{overlapping travel times} that may affect delivery requests in the next time window. Furthermore, \emph{congestion} at recharging nodes may occur if multiple swarms compose the same path simultaneously. Given the aforementioned challenges, \emph{we propose to allocate SDaaS swarms to serve consumers requests optimally from a provider perspective}. This work is the first to introduce the \emph{allocation and re-allocation} problem of homogeneous drone delivery swarms to a set of \emph{time constrained requests} using a \emph{service-oriented approach}. \looseness=-1

We summarize our main contributions as follow:
\begin{itemize}
    \item Define a new set of SDaaS services allocation constraints.
    \item Propose a modified A* congestion-aware algorithm to compose SDaaS services.
    \item Propose three SDaaS allocation algorithms optimizing profit for the providers.
\end{itemize}

\subsection{Motivating Scenario}
\begin{figure}[htbp]
\centerline{\includegraphics[width=3.4in]{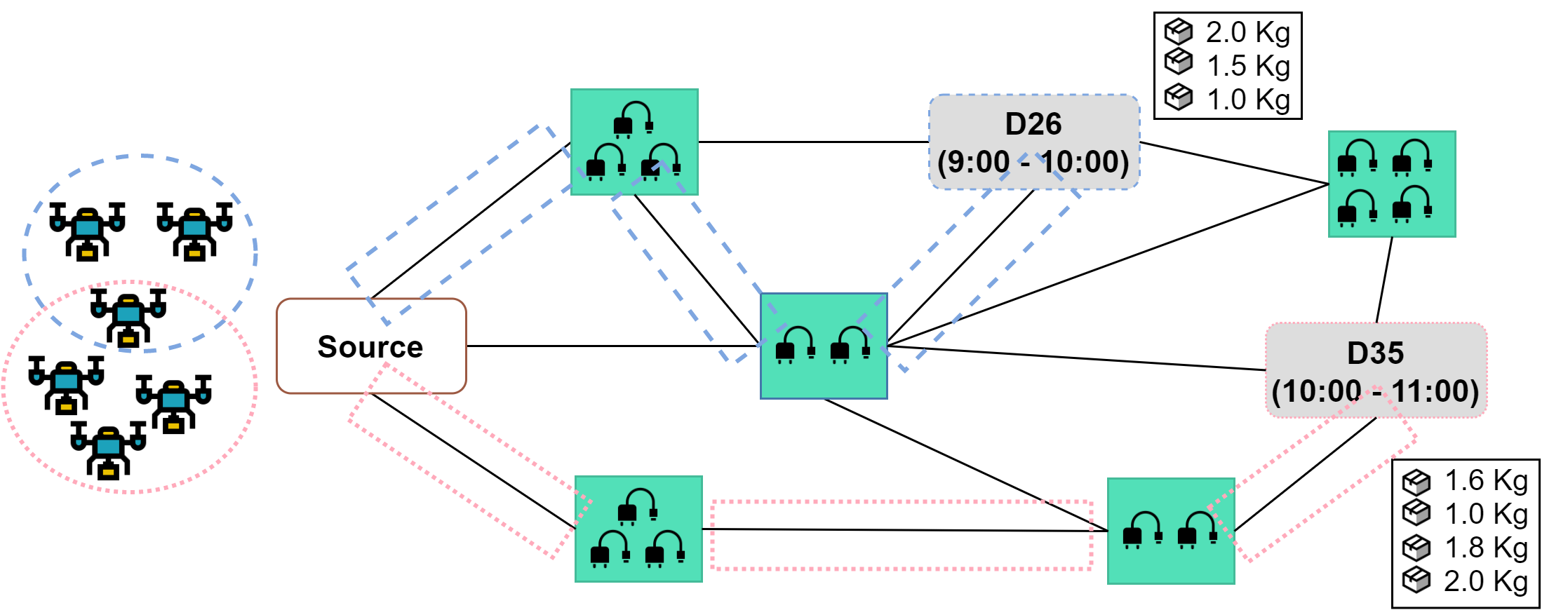}}
\caption{Skyway Network for Multiple Delivery Requests}
\label{motivating}
\end{figure}
Let's assume that, within a day, several medical facilities request multiple packages from a supplier to be delivered together. Hence, multiple swarms are needed to deliver these packages to their respected destinations. Let's also assume that each facility selects a time window for packages arrival. As shown in Fig \ref{motivating}, requested packages may be of different weights. We assume that the maximum package weight does not exceed the drones payload limitations. The weight of the packages directly affects the energy consumption of the drones \cite{alkouz2020swarm}. This in return will affect the distance the swarm can travel before the need to recharge. On the other hand, the medical supplier owns a \emph{finite} set of drones that are used to make different swarms to serve the requests. The drones are \emph{homogeneous}, i.e. same type. Therefore, the drones have same payload capacities, battery capacities, and power consumption rates. These drones would traverse through a line of sight skyway network (following drone fly regulations\footnote{https://www.casa.gov.au/drones/rules/drone-safety-rules}) till they reach the destinations. The intermediate nodes are buildings' rooftops supplied with multiple recharging pads. Each node contains \emph{different numbers of recharging pads}. We assume the environment is deterministic, i.e. we know before hand about the packages weights, availability of recharging pads, battery capacities, etc. As shown in the figure, if the swarm serving destination 26 spends more than an hour to deliver and return back to the source, the delivery to destination 35 will be affected as it needs to re-use a drone from the first swarm to deliver 4 packages.\looseness=-1

Given $r$ requests per day, how to allocate the homogeneous swarms from a finite fleet $n$ to serve the different requests? What is the maximum number of requests could a fleet of size $n$ undertake within a day? How to deal with overlapping travel times? How to allocate swarms to requests that are most profitable for providers? How to re-use and schedule the swarms to maximize the fleet utilization? Furthermore, how to compose the different swarms paths to reduce congestion in the network. \emph{Given all the aforementioned constraints, and without loss of generality, the system should ensure the arrival of the packages to the destinations within consumers' specified time windows.} Future work would extend this work to address environmental issues and uncertainties in the delivery environments.

\section{Related Work}

Swarming can be defined as the collective behavior of a group of entities to achieve a common goal \cite{thornton2018swarming}. Examples in nature include bird flocks, fish shoals, and honey bees \cite{van2008particle}. Similarly, drone swarms can be defined as a set of drones acting as a single entity to achieve a common goal \cite{chen2020toward}. Drone swarms are able to overcome single drones limitations and provide faster and more efficient services \cite{shrit2017new}. In this regard, drone swarms have enormous potential across a broad spectrum of civilian applications. The recent literature spotlights several applications that are already being tested and others envisioned for the future. This includes search and rescue \cite{cardona2019robot}, weather monitoring \cite{hildmann2019nature}, and \emph{delivery} \cite{thornton2018swarming}. 


In delivery, the majority of the literature refer to drone swarms as a set of individually operated drones managed to serve multiple independent deliveries  \cite{san2016delivery}\cite{kuru2019analysis}. However, we define a swarm as a set acting as a single entity and travelling together to deliver multiple packages to a single destination. In this respect, a sequential and parallel drone swarm delivery services compositions was proposed \cite{alkouz2020swarm}. A sequential composition is when a swarm travels together bounded by a space window from a source to a destination. A parallel composition occurs with dynamic swarms capable of splitting and merging between the source and the destination. In addition, the effect of swarm formation and shape on delivery services composition was explored \cite{alkouz2020formation}. However, to the best of our knowledge, no literature discussed the \emph{optimal allocation of drone swarms to serve multiple requests in a delivery context}.

Multi-robot task allocation (MRTA) is an area concerned about the selection of the best robot, out of a set of robots, and allocating it to the best suitable task, out of a set of tasks, to optimize an overall system performance \cite{khamis2015multi}. This area also deals with homogeneous and heterogeneous robots. Homogeneous robots are robots within a group with same capabilities \cite{gigliotta2018equal}. Heterogeneous robots are robots within a group with different capabilities \cite{dorigo2013swarmanoid}. For example, homogeneous drones are drones with same speeds, sizes, or payload and battery capacities. The tasks that require multiple robots to be executed are referred to as Multi-robot Tasks \cite{khamis2015multi}. In this regard, to the best of our knowledge, the allocation of homogeneous drone swarms to a set of \emph{constrained} tasks has not yet been explored. In addition, the re-allocation and scheduling of swarms to perform multiple tasks has not been addressed. Thus, this work is the first to introduce the \emph{allocation and re-allocation} problem of homogeneous swarms (drone delivery swarms) to a set of time \emph{constrained tasks} (consumers requests) using a \emph{service-oriented approach}.

The service paradigm leverages the drone technology to effectively provision drone-based services \cite{shahzaad2021resilient}. The service paradigm presents a higher level of abstraction that provides congruous solutions to real-world problems \cite{bouguettaya2017service}. Single drone delivery services framework was presented as Drone-as-a-Service (DaaS) \cite{shahzaad2020game}. This framework was later expanded to model swarm-based services called Swarm-based Drone-as-a-Service (SDaaS) as they map to the key ingredients of the service paradigm, i.e., functional and non-functional attributes \cite{alkouz2020swarm}. Hence, this work \emph{extends} the SDaaS framework to cover the key element of \emph{allocating} the service swarm members to optimize the service's QoS from a \emph{provider perspective}. This work also considers \emph{SDaaS services composition} and delivery delays that may be caused due to congestion and the concurrent use of the skyway network medium.

\section{Swarm-based Drone-as-a-Service Model}

In this section, a service model for swarm-based delivery services
is presented. \emph{We abstract each swarm travelling on a skyway between two nodes as an SDaaS} (see Fig. \ref{motivating}). We formally define a Swarm-based Drone-as-a-Service (SDaaS). We then define an SDaaS provider and a consumer's request. Later, we discuss the constraints surrounding the allocation of SDaaS services.\\
\textbf{Definition 1: Swarm-based Drone-as-a-a-Service (SDaaS).} An SDaaS is defined as a set of drones, i.e. more than one, carrying packages and travelling in a skyway segment. It's represented as a tuple of $<SDaaS\_id, S, F>$, where
\begin{itemize}
    \item $SDaaS\_id$ is a unique service identifier
    \item $S$ is the swarm travelling in SDaaS. $S$ consists of $D$ which is the set of drones forming $S$, a tuple of $D$ is presented as $<d_1,d_2,..,d_m>$. $S$ also contains the properties including the current battery levels of every $d$ in $D$ $<b_1,b_2, ..,b_m>$, the payloads every $d$ in $D$ is carrying $<p_1,p_2,..,p_m>$, and the current node $N$ the swarm S is at.
    \item F describes the delivery function of a swarm on a skyway segment between two nodes, A and B. F consists of the segment distance $dist$, travel time $tt$,  charging time $ct$, and waiting time $wt$ when recharging pads are not enough to serve $D$ simultaneously in node B.
\end{itemize}
\textbf{Definition 2: SDaaS Service Provider.} A provider is presented as a tuple of $<D, \alpha>$, where
\begin{itemize}
    \item $D$ is the finite set of $n$ drones owned by the provider. $D$ is a tuple of $<d_1,d_2,..,d_n>$ and every drone $d_i$ consists of a tuple of $<b,p,s>$ where $b$ is the maximum battery capacity of the drone, $b$ is the maximum payload capacity of the drone, and $s$ is the maximum speed of the drone.
    \item $\alpha$ is the providers locations, i.e. source node.
\end{itemize}
\textbf{Definition 3: SDaaS Request.} A request is a tuple of $< R\_id,\beta, P, T>$, where
\begin{itemize}
    \item $R\_id$ is the request unique identifier.
    \item $\beta$ is the request destination node.
    \item $P$ are the weights of the packages requested, where $P$ is $<p_1,p_2,..p_m>$.
    \item $T$ is the time window of the expected delivery, it is represented as a tuple of $<st,et>$ where $st$ is the start time of the requested delivery window and $et$ is the end time of the requested delivery window.
\end{itemize}

\subsection{SDaaS members allocation constraints}
\begin{figure}[htbp]
\centerline{\includegraphics[width=3.5in]{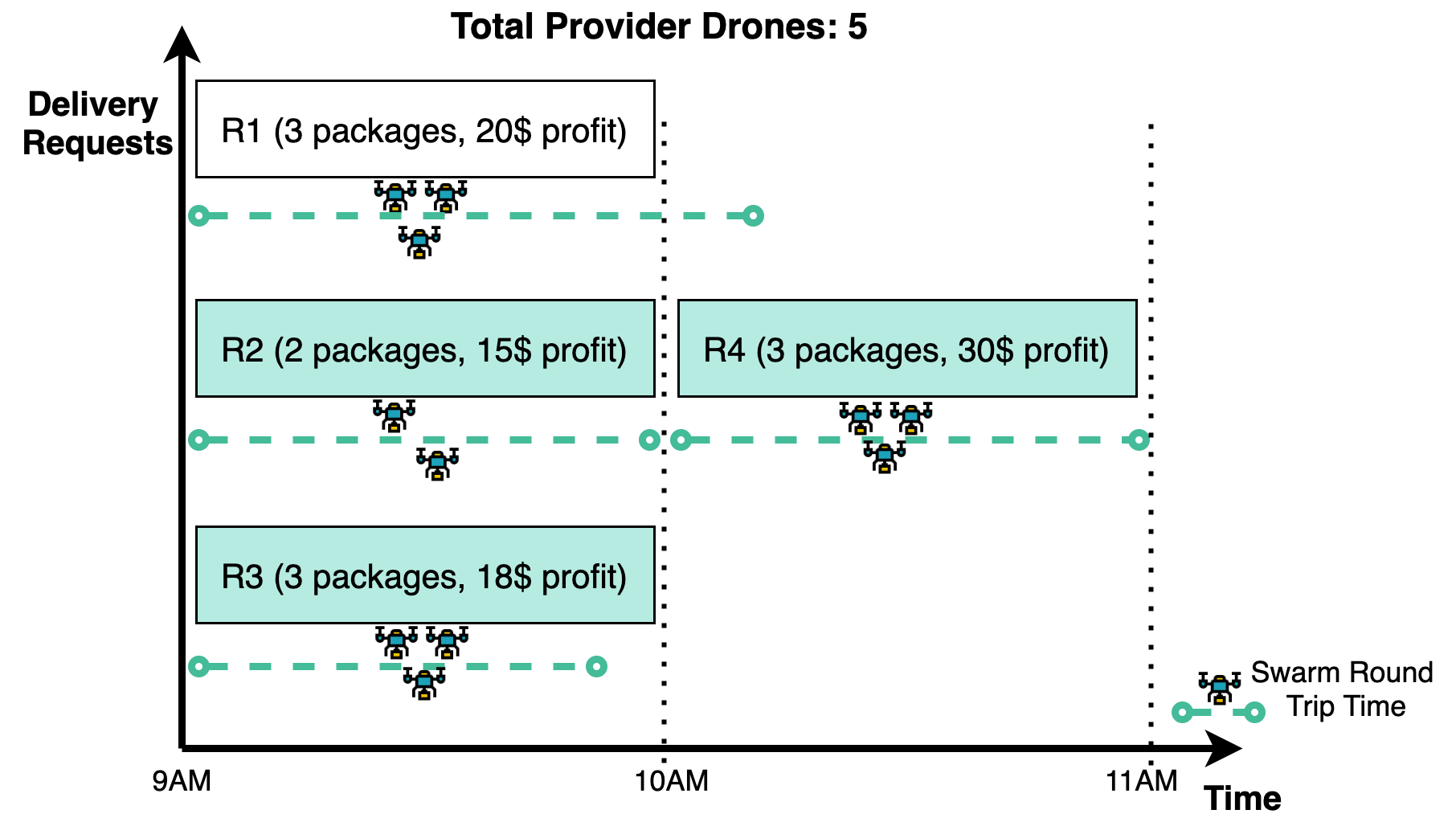}}
\caption{Allocation and Scheduling Constraints in SDaaS}
\label{constraints fig}
\end{figure}
We address the following constraints surrounding the swarm-based services allocation to deliver on its potential:
\begin{itemize}[leftmargin=*]
    \item \emph{Strict time of delivery:} A delivery request is strictly bounded by a time window of packages arrival. This means if a package arrives before the window start time or after the window end time the delivery is considered unsuccessful. If a package arrives earlier than scheduled, a consumer might not be at the destination location to receive it. In a similar manner, if a package arrives late, a consumer will not be satisfied by the service.
    \item \emph{Time overlapped requests:} There might be instances where multiple consumers requests are to be delivered within the same time window. Since the provider has a finite number of $n$ drones, some overlapped requests may not be served. This is true if the number of packages exceed the number of available drones. In this case, the provider needs to select the most profitable requests and reject the least profitable ones. Fig.\ref{constraints fig} depicts this constraint were R1-R3 are overlapping and only two requests could be served at a time since the provider owns 5 drones only.
    \item \emph{Inter-dependency of delivery requests:} The finite drones need to be allocated to the requests that will maximize the profit for the provider. Hence, the drones may be allocated multiple times to serve multiple requests. However, there might be instances when selecting a most profitable request within a time window may lead to a non-optimal profit (R1 in Fig. \ref{constraints fig}). This is because the long Round Trip Time (RTT) for the R1 swarm will lead to opting out R4 that is more optimal. In addition, selecting R4 allows the selection of R3 which would not have been selected if R1 is selected. Hence, the allocation should consider the time inter-dependent requests to maximize the provider profit throughout the full day.
    \item \emph{Skyway network contentions:} Congestion at nodes may occur since concurrent requests mean that multiple swarms will be using the skyway network at the same time. This will cause increased delivery times. Hence, we need to address the SDaaS service composition and path selection to reduce the number of sequential charges due to congestion.
    \item \emph{SDaaS environment constraints:} In addition to all the aforementioned constraints, the authors of \cite{alkouz2020swarm} has defined some constraints in the SDaaS environment. This includes the different battery consumption rates within a swarm due to different carried payloads. In addition, the number of drones in a swarm may exceed the number of available recharging pads at a node causing sequential charging.
\end{itemize}

Our aim is to successfully allocate and re-allocate swarms to serve the best set of consumers requests, from a provider's perspective, given the aforementioned constraints. To the best of our knowledge, this paper is the first to propose the optimal allocation of swarm services to consumers' requests.

\section{SDaaS Members Selection and Allocation Framework}
\begin{figure*}[t]
\centerline{\includegraphics[width=\linewidth]{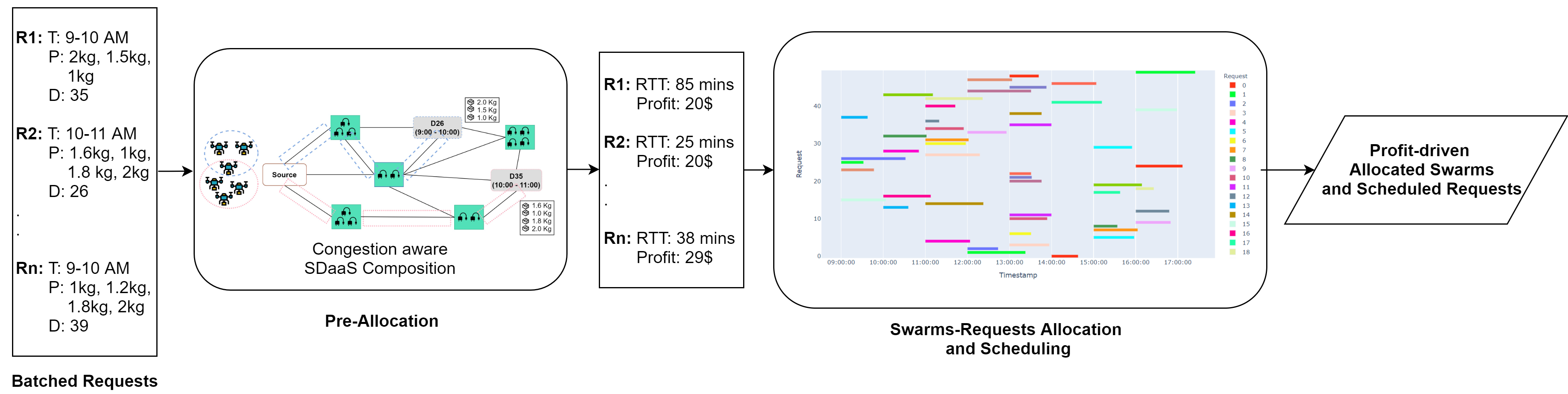}}
\caption{SDaaS Members Allocation Framework}
\label{framweork_fig}
\end{figure*}
The SDaaS members selection and allocation framework is a provider-centric, profit-driven, framework. The framework consists of two main modules as shown in Fig.\ref{framweork_fig}. The first is the congestion-aware SDaaS services composition (path composition) for every request. The output of this module is the maximum Round Trip Time (RTT) for every request and the profit if the request is allocated and served. The second module uses the maximum RTT produced by module one and allocates the drones to the requests to maximize the utilization of the provider's drones and increase the providers profit.
\subsection{Congestion-aware SDaaS Composition}
This paper focuses on a provider owning a \emph{finite} set of \emph{homogeneous} drones, i.e. drones with same capabilities. We assume that the packages in a request do not exceed the payload capacities of the drones. In addition, a drone can carry one package at a time and serve one request in a single trip. Therefore, the number of drones in a swarm, serving a request, is equal to the number of packages in a request. We also assume a request may consist of maximum $m$ packages, i.e. maximum $m$ drones form a swarm. The goal of this module is to compute the maximum time taken by a trip taken by a swarm to serve a request and return back to the origin $RTT$. Computing the $RTT$ for a request is important in order to reuse the drones in other requests. The $RTT$ is highly affected by the path taken by the swarm to the destination and its return path. The number of recharging pads, as described earlier, is \emph{different at each node}. Some nodes may have recharging pads that are less than the number of drones in the swarm. Hence, sequential charging may occur adding to the $RTT$. In addition, contention may occur at a node if two swarms serving different requests take the same path at a time causing \emph{congestion}. Hence, we propose a modified A* \emph{congestion-aware} SDaaS services composition algorithm. The $RTT$ along with the number of drones used determine the profit of the provider if the request is served. We assume that the environment is deterministic, i.e. we know the availability of recharging pads considering other providers using the network.
\begin{algorithm}
 \caption{Congestion-aware Services Composition Algorithm}
 \label{congestionlAlg}
 \small
 \begin{algorithmic}[1]
 \renewcommand{\algorithmicrequire}{\textbf{Input:}}
 \renewcommand{\algorithmicensure}{\textbf{Output:}}
 \REQUIRE $S$, $R$
 \ENSURE  $RTT$, Profit
 \STATE $RTT$ = 0
    \WHILE{$S$ is not at destination and back at the source}
        \STATE distance to destination= \textbf{Dijkstra}(current, destination)
        \STATE \textbf{compute} energy consumption for every $d$ in $S$ based on $R$ package weights and distance to destination
        \IF{all $d$ in $S$ can reach destination without intermediate nodes}
        \STATE $S$ travels to destination
        \STATE $t$+=travel time
        \ELSE
        \STATE \textbf{find} all nearest neighbor nodes
            \IF {$P_D-S_D < m$}
            \STATE available\_pads=total\_pads - ($P_D-S_D$)
            \ELSE
            \STATE available\_pads=total\_pads - $m$
            \ENDIF
        \STATE \textbf{select} best neighboring node (min travel time and min node time)
        \STATE $S$ travels to neighboring node
        \STATE $RTT$+=travel time + charging time + waiting time
        \ENDIF
    \ENDWHILE
    \STATE Profit = \textbf{size($S_D$)} * RTT * Constant
 \RETURN $RTT$, Profit
 \end{algorithmic}
 \end{algorithm}
All the drones ($S_D$) serving a request, fully charged, form a swarm at the source node and traverse the network without dispersing in a static behaviour \cite{akram2017security}. While the drone is not at the destination, the swarm computes the potential to reach the destination directly without stopping at intermediate nodes to recharge using the shortest Dijkstra's path \cite{dijkstra1959note}. The potential is computed by considering the payload of all drones and its effect on the battery consumption. If the swarm can travel directly, the $RTT$ is updated by the travel time $tt$ in the selected segments. If the swarm can't reach the destination directly, it greedily selects the most optimal neighbouring node. The most optimal neighbour is the one with the least travel time $tt$, charging time $ct$, and waiting times $wt$ caused by sequential charging due to limited number of recharging pads. Since the payloads carried are different between the drones, the energy consumption rate is different, and the battery charging times are different $ct$. We take the maximum charging time of concurrent recharging drones and add it to the $RTT$. At every node, we consider the \emph{potential of congestion} to compute the maximum possible $RTT$ that will \emph{guide the allocation process}. We assume that each node is occupied by all the other drones owned by a provider if  they are less than the maximum swarm size $P_D-S_D< m$. Otherwise, we consider the node to be occupied by a swarm of maximum size $m$. We compute the node time, i.e. $ct+wt$, considering the number of \emph{available recharging pads} under congestion. We assume no more than two swarms may select a node at a time. The $RTT$ is updated and the drone again checks its potential to reach the destination. The process repeats until the swarm is at the destination. Once at the destination, the swarm charges full and travels back to the source in the same manner. The energy consumption rate at the return trip is much lower than the outbound trip due to the released payload. The $RTT$ is updated at the source after charging fully. Once the $RTT$ is computed the profit of the request, if allocated and served, is computed using the $RTT$ and the number of drones serving the request $S_D$. The profit is computed per mile per drone taking the constraints into consideration. Algorithm \ref{congestionlAlg} describes the congestion-aware services composition process.

\subsection{SDaaS members Allocation and Requests Scheduling}


The output of the first module, i.e. the maximum $RTT$ and Profit, for all requests is used to allocate the drones swarm members to the most profitable combination of requests. In this section, we propose three methods for the swarm-request allocation. We assume that the requests are received in batch for the full day. We divide the day into $t$ time windows $W$. For simplicity, we assume that a drone may be used once in a time window. Hence, if a request's $RTT$ overlaps over two time windows, we assume that the swarm serving this request is booked for both time windows. This assumption would be relaxed in the future work. We need to allocate the swarms to serve the most profitable requests while maximizing the utilization of the \emph{finite} number of drones that the provider owns.\looseness=-1

The swarms-requests allocation problem is a part of the Multi-Robot tasks allocation problems. However, to the best of our knowledge, no previous work has addressed the allocation problem from a service provider perspective to allocate a \emph{finite} set of robots to \emph{time-constrained} consumers requests (tasks) and ensuring the \emph{re-use} of robots to maximize their utilization. Each drone may serve more than one request at different times. We propose to tackle this problem through three proposed algorithms namely: Time-based greedy algorithm, Request-based greedy algorithm, and the Heuristic-based algorithm. Later, we will compare the performance of these algorithms against the Brute force algorithm.

\subsubsection{Request-based greedy allocation algorithm}
In this approach, all the requests are first sorted descendingly by their profit. The algorithm allocates swarms to the requests in order of profit if the request is servable with the available drones. Algorithm \ref{request-based} describes the request-based greedy algorithm. $P_D$ is the finite number of drones owned by the provider. The requests are looped through once. Each request checks for \emph{two conditions}. The \emph{first condition} checks if there are available drones within its window ($W_{R_i}$) to serve the request [line 4]. This is done by checking if the number of the request's drones ($D_{R_i}$) plus the number of used (occupied) drones ($used\_drones[W_{R_i}]$), within the same time window, are less than or equal to the total owned drones of the provider ($P\_D$). The \emph{second condition} is checked if the first condition is true. Here, the algorithm checks if the request's $RTT$ overlaps over the next time window ($W_{R_i}+1$) [line 5]. In this case, if there are available drones in the next time window the swarm gets allocated to serve the request ($served\_R$). The allocated request's profit ($P_{R_i}$) is added to the $total\_profit$, the request's drones ($D_{R_i}$) are added to the used drones of the two time windows ($used\_drones[W_{R_i},W_{R_i}+1]$) and to the total utilized drones $drones\_utilized$. Otherwise, if there are no available drones in the next time window $W_{R_i}+1$ the request is discarded. If the second condition is not satisfied [line 12], i.e. the request does not overlap over the next time window, the swarm is allocated to serve the request, and are added to the $used\_drones [W_{R_i}]$ of the first time window only. The process continues until all requests are looped through. The sum of the profits of the served requests and the drones used in them is returned at the end.
\begin{algorithm}
 \caption{Request-based greedy allocation algorithm}
 \label{request-based}
 \small
 \begin{algorithmic}[1]
 \renewcommand{\algorithmicrequire}{\textbf{Input:}}
 \renewcommand{\algorithmicensure}{\textbf{Output:}}
 \REQUIRE $R$, $P\_D$
 \ENSURE  served\_R, total\_profit, drones\_utilized 
 \STATE total\_profit = 0 , drones\_utilized = 0 ,
 \\ served\_R = $[\:]$ , used\_drones = $[\:]$ 
 \STATE sorted\_R = \textbf{sort\_by} (R, profit)
 \FOR{\textbf{each} R \textbf{in} sorted\_R}
    \IF{$(D_{R_i} + $ used\_drones$[W_{R_i}]) <= P\_D$}
        \IF{\textbf{overlap} ($RTT_{R_i},W_{R_i}+1$) == True }
        \IF{$(D_{R_i} + $ used\_drones$[W_{R_i}+1]) <= $  $P\_D$}
            \STATE served\_R.\textbf{append}($R_i$)
            \STATE total\_profit += $P_{R_i}$ 
            \STATE used\_drones$[W_{R_i},W_{R_i}+1]$ += $D_{R_i}$
            \STATE drones\_utilized += $D_{R_i}$
        \ENDIF
        \ELSE 
            \STATE total\_profit += $P_{R_i}$ 
            \STATE used\_drones$[W_{R_i}]$ += $D_{R_i}$ 
            \STATE served\_R.\textbf{append}($R_i$)
            \STATE drones\_utilized += $D_{R_i}$
        \ENDIF
    \ENDIF
 \ENDFOR
 \RETURN (served\_R, total\_profit, drones\_utilized)
 \end{algorithmic}
 \end{algorithm}

\subsubsection{Time-based greedy allocation algorithm}
This approach is a slight modification to the request-based greedy algorithm where the request time window is taken in consideration when sorting the requests. Here, all the requests are first sorted by their start times ascendingly. All the requests within a time window are then sorted by their profit descendingly. The rest of the algorithm checks the validity of the requests combination similar to Algorithm \ref{request-based} and computes the total profit, drones utilized, and the set of the served and allocated requests.

\subsubsection{Heuristic-based allocation algorithm}
The main limitation of the greedy algorithms is the order of the requests processed and served. Sorting the requests on profit and time does not ensure the most optimal solution. This approach is, hence, a modification of the greedy algorithms that tries to tackle the order limitation. Algorithm \ref{Heuristic-based} depicts the heuristic-based allocation method. In this approach, the unsorted list of requests are looped through considering different allocation start points resulting in different combinations of requests. Two main set of arrays and variables are constructed. The first set of arrays are of the size of the requests $R$ storing the profits ($all\_profits$), drones utilized ($all\_drones\_utilized$), and the combination of served requests ($all\_served\_R$) generated by different starting points (starting request). The second set of arrays and variables are local temporary active arrays that are re-constructed at every starting point (request). These arrays and variables store the total profit ($active\_total\_profit$), drones utilized ($active\_drones\_utilized$), and the served requests combination ($active\_served\_R$) with a certain starting point (request). A nested loop is constructed where the outer loop loops over all requests $R$ [line 2]. The inner loop, similar to the greedy, loops over the requests and checks the valid combination \emph{however starting from the request} which the outer loop is at ($r\_id$) [line 4]. After every iteration of the inner loop, the variables of the valid served requests are updated in the active set of arrays and variables. After the full iteration of the inner loop, \emph{the first set of arrays}, of size $R$, store the total profit, drones utilized, and the combination of served requests at the index of the starting point (request) ($r\_id$). Once the outer loop finishes, the algorithm finds the combination of requests with the maximum profit and the number of drones utilized in this combination [lines 25-28].

\begin{algorithm}
 \caption{Heuristic-based allocation algorithm}
 \label{Heuristic-based}
 \small
 \begin{algorithmic}[1]
 \renewcommand{\algorithmicrequire}{\textbf{Input:}}
 \renewcommand{\algorithmicensure}{\textbf{Output:}}
 \REQUIRE $R$, $P\_D$
 \ENSURE  served\_R, total\_profit, drones\_utilized 
 \STATE all\_profits = [\:] , all\_drones\_utilized = [\:] ,
 \\ all\_served\_R = $[\:]$  
 \FOR{\textbf{each} r \textbf{in} R}
    \STATE active\_total\_profit = 0 , active\_served\_R = [\:] , 
    \\ active\_used\_drones = [\:] ,  active\_drones\_utilized = 0
    \FOR{ \textbf{each} AR in \textbf{loop\_around} ($r\_id$)}
    \IF{$(D_{AR_i} + $ used\_drones$[W_{AR_i}]) <= P\_D$}
        \IF{\textbf{overlap} ($RTT_{AR_i},W_{AR_i}+1$) == True }
        \IF{$(D_{AR_i} + $ used\_drones$[W_{AR_i}+1]) <= $  $P\_D$}
            \STATE active\_served\_R.\textbf{append}($AR_i$)
            \STATE active\_total\_profit += $P_{AR_i}$
            \STATE active\_used\_drones$[W_{AR_i},{W_{AR_i}+1}]$ += $D_{AR_i}$
            \STATE active\_drones\_utilized += $D_{AR_i}$
        \ENDIF
        \ELSE 
            \STATE active\_total\_profit += $P_{AR_i}$
            \STATE active\_used\_drones$[W_{AR_i}]$ += $D_{AR_i}$ 
            \STATE active\_served\_R.\textbf{append}($AR_i$)
            \STATE active\_drones\_utilized += $D_{AR_i}$
        \ENDIF
    \ENDIF
    \ENDFOR
 \STATE all\_profits.\textbf{append}(active\_total\_profit)
 \STATE all\_drones\_utilized.\textbf{append}(active\_drones\_utilized)
 \STATE all\_served\_R.\textbf{append}(active\_served\_R) 
 \ENDFOR
 \STATE total\_profit= \textbf{max}(all\_profits)
 \STATE max\_index =all\_profits.\textbf{index}(total\_profit)
 \STATE served\_R = all\_served\_R[max\_index]
 \STATE drones\_utilized = all\_drones\_utilized[max\_index]
 \RETURN (served\_R, total\_profit, drones\_utilized)
 \end{algorithmic}
 \end{algorithm}

\begin{figure}[htbp]
\centerline{\includegraphics[width=3.6in]{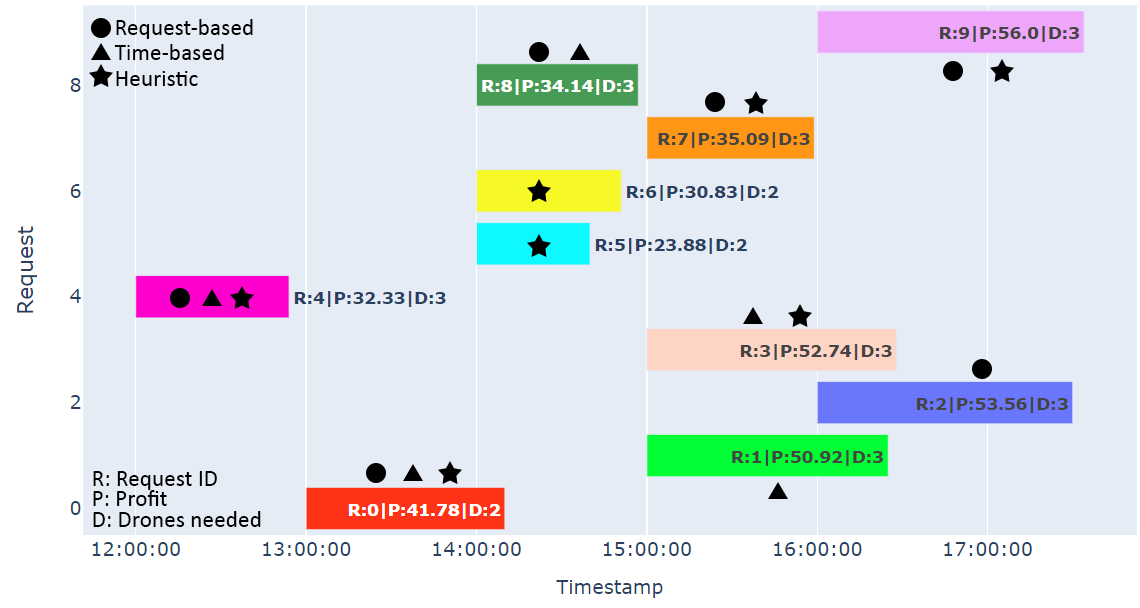}}
\caption{Requests allocation with proposed methods (provider owned drones=6)}
\label{allocation}
\end{figure}

Fig. \ref{allocation} shows an allocation example of the three proposed algorithms with a set of 10 requests (0-9). The bars represent the maximum $RTT$ for each request. In this example, we assume that the provider owns 6 drones only. The drones are used to make up swarms to serve the different requests. As mentioned earlier, we assume that a drone once used in a time window may not be used again within the same window. The algorithms result in a set of selected requests that the drones get allocated to. As shown in the figure, the heuristic-based algorithm outperforms the two methods with a total profit of 272.65 and 7 served requests out of 10. Next, the request-based greedy algorithm served 6 requests with a total profit of 252.9. Finally, the time-based greedy algorithm served 5 requests with a total profit of 211.91. 

The main down-side of the \emph{time-based greedy algorithm} is the starting time window. If requests are served and allocated in time window $i$ use all/most drones and the served requests overlap to the next time window ($i+1$), there are chances that more profitable requests in the next time window will not be served. As shown in Fig.\ref{allocation}, since requests 1 and 3 have been allocated, as they arrive earlier, the more profitable requests 2 and 9 were not served. In the same manner, the \emph{request-based greedy algorithm} chooses the most profitable request but there might be a combination of other smaller requests with a higher profit. As shown in the figure, the algorithm selects request 8 to be served whereas the total profit of requests 5 and 6 is higher. Since the \emph{heuristic-based algorithm} overcomes the limitation of requests order, it smartly outperforms the two other methods and allocates swarms to serve requests that might be less profitable but in total resulting in more profit and more served requests.

\section{Experiments and Results}

\begin{table}[t]
\caption{Experiment Variables}
\label{tab:variables}
\begin{tabular}{l|l}
\hline
Variable                                                                                     & Value                                                                                                                                        \\ \hline
PC information                                                                               & {\color[HTML]{333333} \begin{tabular}[c]{@{}l@{}}7th Gen Intel® Core™ i7-\\ 7700HQ Processor ( 2.8 GHz)\\ 16 GB RAM, 64-bit OS\end{tabular}} \\
\begin{tabular}[c]{@{}l@{}}No. of nodes in the largest \\ connected sub-network\end{tabular} & 195                                                                                                                                          \\
Max no. of packages in a request                                                             & 5                                                                                                                                            \\
Max weight of a package                                                                      & 1.4 kg                                                                                                                                       \\
Drone model                                                                                  & DJI Phantom 3                                                                                                                                \\
Time for drone to fully charge                                                               & 30 mins                                                                                                                                      \\
Battery capacity                                                                             & 4480 mAh                                                                                                                                     \\
Drone speed                                                                                  & 15.6 m/s                                                                                                                                     \\ \hline
\end{tabular}
\end{table}

In this section, we evaluate the performance of the three proposed methods of allocations. We conduct a set of experiments to evaluate the performance in terms of profit maximization, drones utilization, requests fulfilment, and computation. Although there are existing resource allocation algorithms \cite{carlier1989algorithm}\cite{salkin1975knapsack}, the proposed problem is fundamentally different in that it is machine (drone) unspecific, re-allocatable, and time dependant. Hence, we compare the proposed methods against the \emph{allocation-optimal} and \emph{exhaustive} Brute force allocation to measure their performance. 

The dataset used in the experiments is an urban road network dataset from the city of London. The data consist of a graph edge list of the city \cite{karduni2016protocol}. For the experiments we took a sub-network of 129 connected nodes to mimic a skyway network. Each node was allocated with different number of recharging pads randomly. We then set a source node and generate $r$ requests with different destination nodes. For each request, we synthesize payloads with a maximum size of 5 packages and a maximum weight of 1.4 kg. We use the congestion-aware SDaaS composition algorithm to compute the  profit and the maximum $RTT$ of each request given the different number of recharging pads at each node. Each generated request is randomly assigned to a delivery time window. We assume the drone takes 30 minutes to fully charge and travel time is computed using the distance of each segment in the composed path. The rest of the experimental variables are presented in table \ref{tab:variables}.

\begin{figure}[t]
\centerline{\includegraphics[width=3.5in]{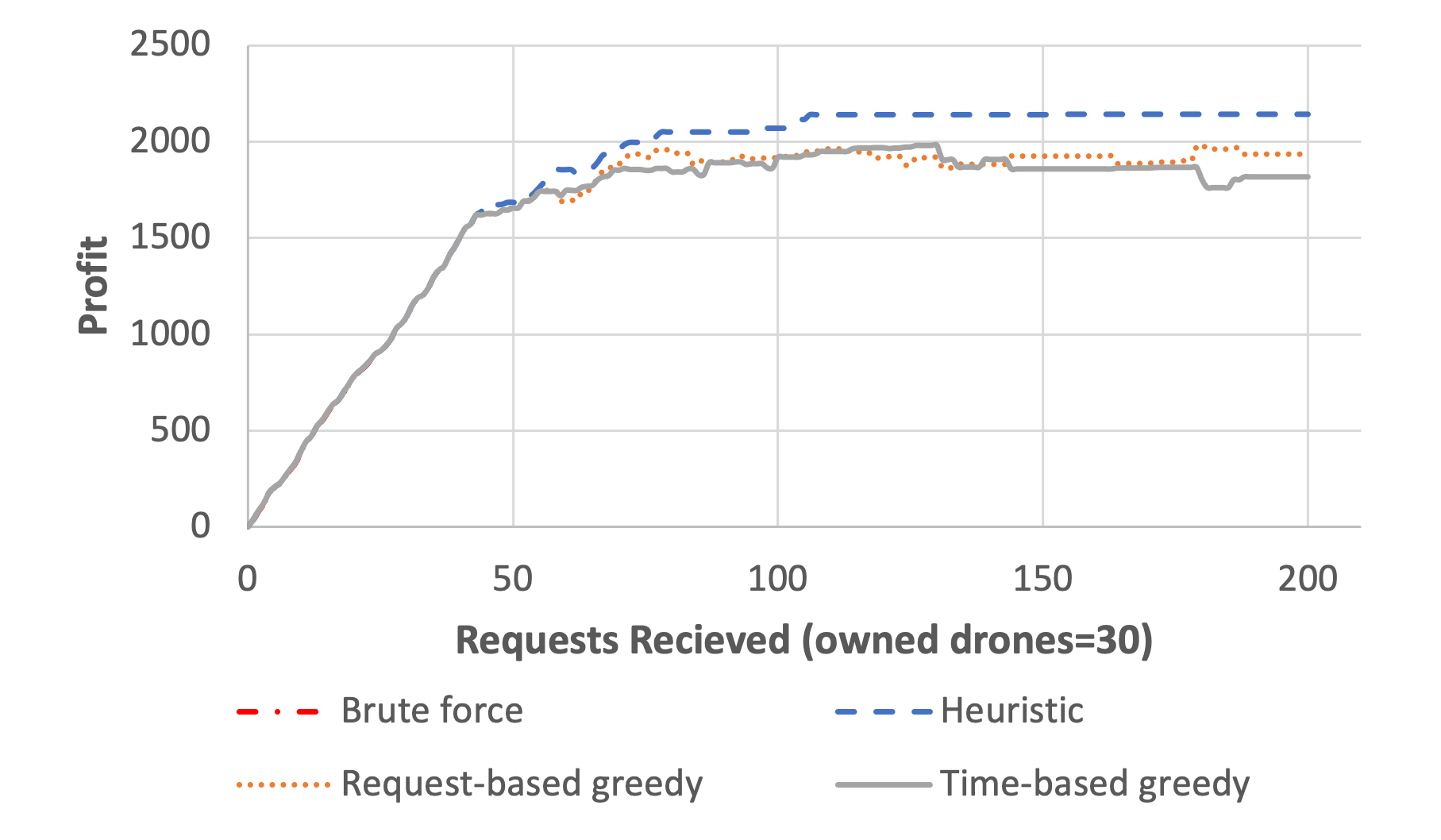}}
\caption{Profit with varying number of received requests}
\label{profit-requests}
\end{figure}

In the first experiment, we compare the performance in terms of profit of the three proposed algorithms against the \emph{optimal} and \emph{exhaustive} Brute force allocation approach. In the brute force approach, all possible combination of requests are generated. Combinations that are non feasible due to the finite number of provider's drones are discarded. The feasible combination with the highest profit is returned and the swarms are allocated to the requests of this combination. In this experiment, we assume a provider owns 30 drones that are allocated and re-used. As shown in Fig. \ref{profit-requests}, the heuristic-based approach outperforms the request-based and time-based greedy algorithms. The x-axis represents the number of requests received within a day. The y-axis represents the total profit made by serving the allocated requests. The brute force approach, as shown in the figure, terminated at 27 requests. This is because the memory usage is highly exponential as all possible combination of requests are generated and processed. This shows that the brute force approach is not feasible in real-word scenarios where the number of delivery requests received a day may be large in number. The heuristic-based approach outperforms the request-based and time-based greedy algorithms. As explained earlier, this is due to the heuristic-based algorithm capability of overcoming the order of allocation constraint (it does not allocate the most profitable requests first as shown in Fig. \ref{allocation}). The heuristic-based algorithm smartly outperforms the two other proposed methods and allocates swarms to serve requests with less profit but in total resulting in more profit and more served requests. The request-based and time-based greedy algorithms are interchangeably preforming better than each other. This, as described earlier, mainly depends on the order of the received requests. In some cases, where requests with higher profits are widely distributed over the time windows, the time-based method would perform better. Otherwise, if most profitable requests are not distributed, the request-based method performs better. 

\begin{figure}[t]
\centerline{\includegraphics[width=3.5in]{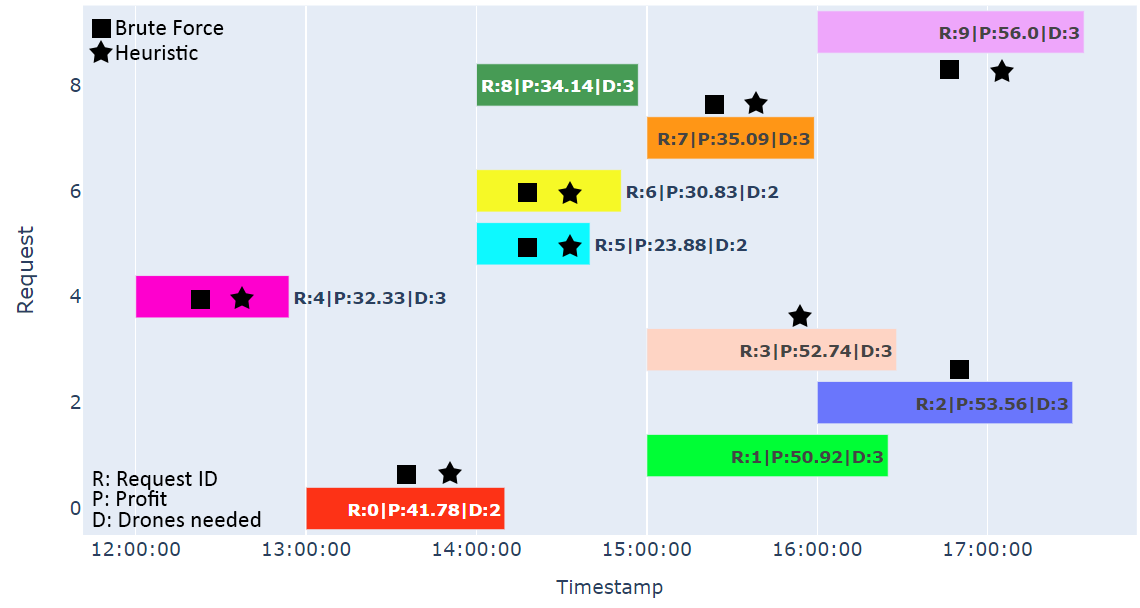}}
\caption{Requests allocation of Brute Force Vs Heuristic (provider owned drones=6)}
\label{brute-heuristic}
\end{figure}

\begin{figure}[t]
\centerline{\includegraphics[width=3.5in]{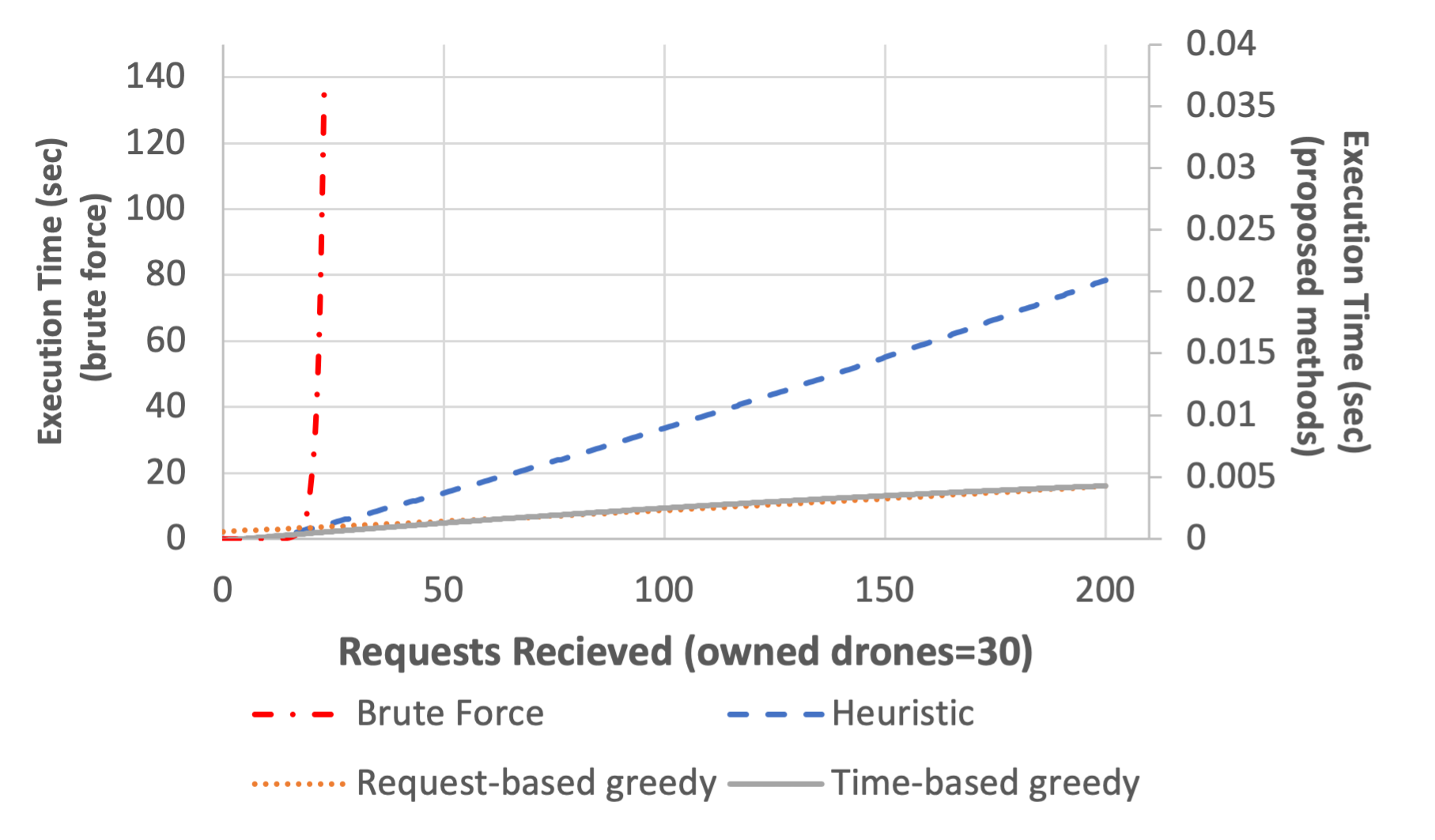}}
\caption{Execution times of proposed methods}
\label{execution}
\end{figure}

Although the heuristic-based method is not always optimal, it is the best performing \emph{feasible} method. The heuristic-based method in few cases does not give the optimal solution when compared to the brute force approach. Fig. \ref{brute-heuristic} shows an example of such situation. As shown in the figure, the only difference between the heuristic and brute force is the allocation of requests 2 (more profitable) and 3. Since the heuristic loops over the requests in sequence, changing the starting point at every iteration, there are chances that requests with lower profits get allocated and requests with higher profits coming later do not find drones to be served. The selected requests in the figure started at iteration 3, i.e. request 3 was the first to get allocated. Since requests 3, 7, and 9 got allocated (in sequence) with the heuristic method, requests 1 and 2 did not have enough drones in their respected time windows to get allocated. Regardless of the limitation described, the execution time of the heuristic-based algorithm is polynomial compared to the exponential time of the brute force making it a more scalable and feasible solution. Fig. \ref{execution} shows the execution times of all methods varying the number of requests received a day. The left y-axis represents the execution time for the brute force algorithm. The right y-axis represents the execution times of the three proposed methods.

\begin{figure}[t]
\centerline{\includegraphics[width=3.5in]{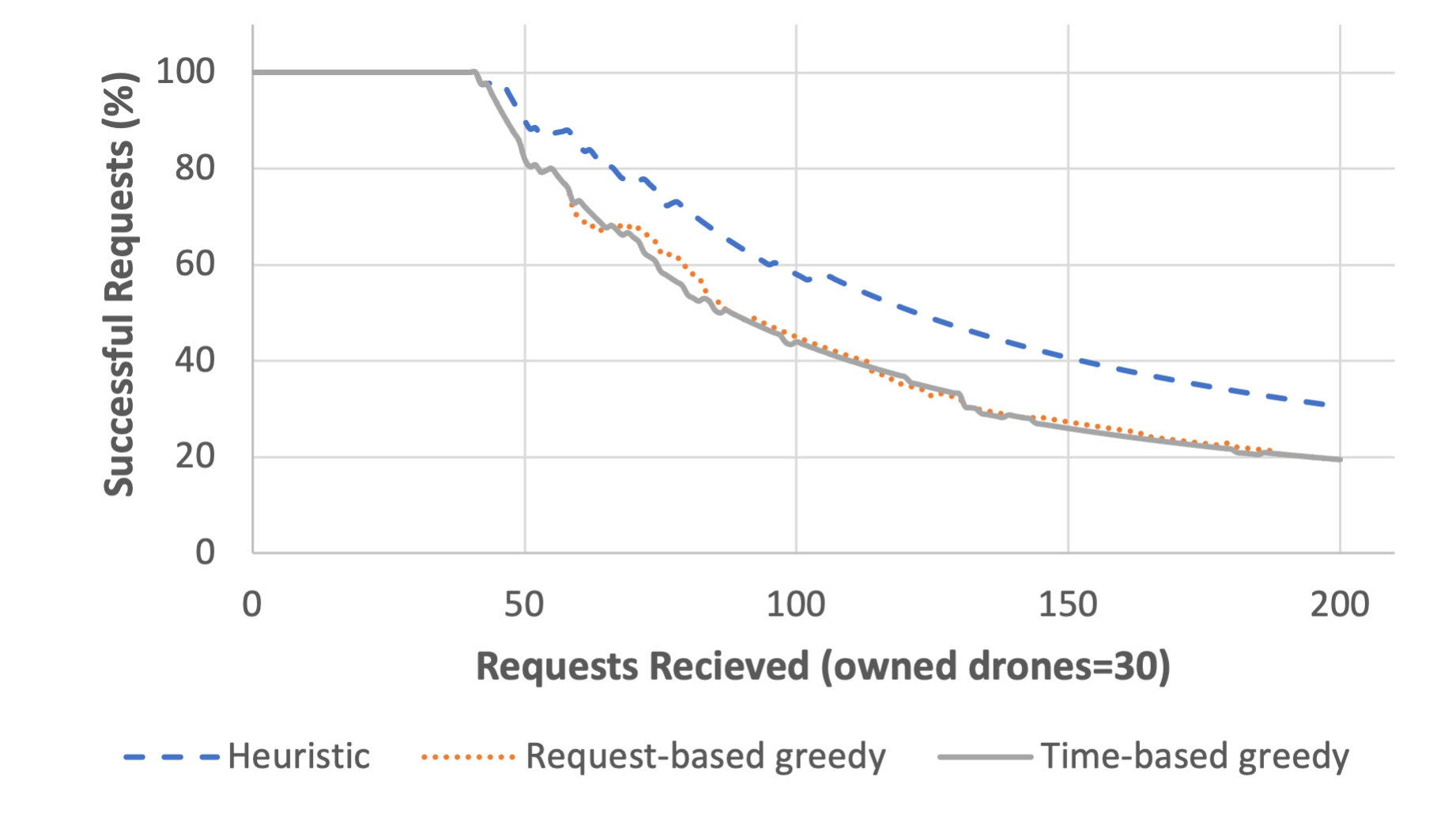}}
\caption{Successful requests varying the number of received requests}
\label{successful-requests}
\end{figure}

\begin{figure}[t]
\centerline{\includegraphics[width=3.5in]{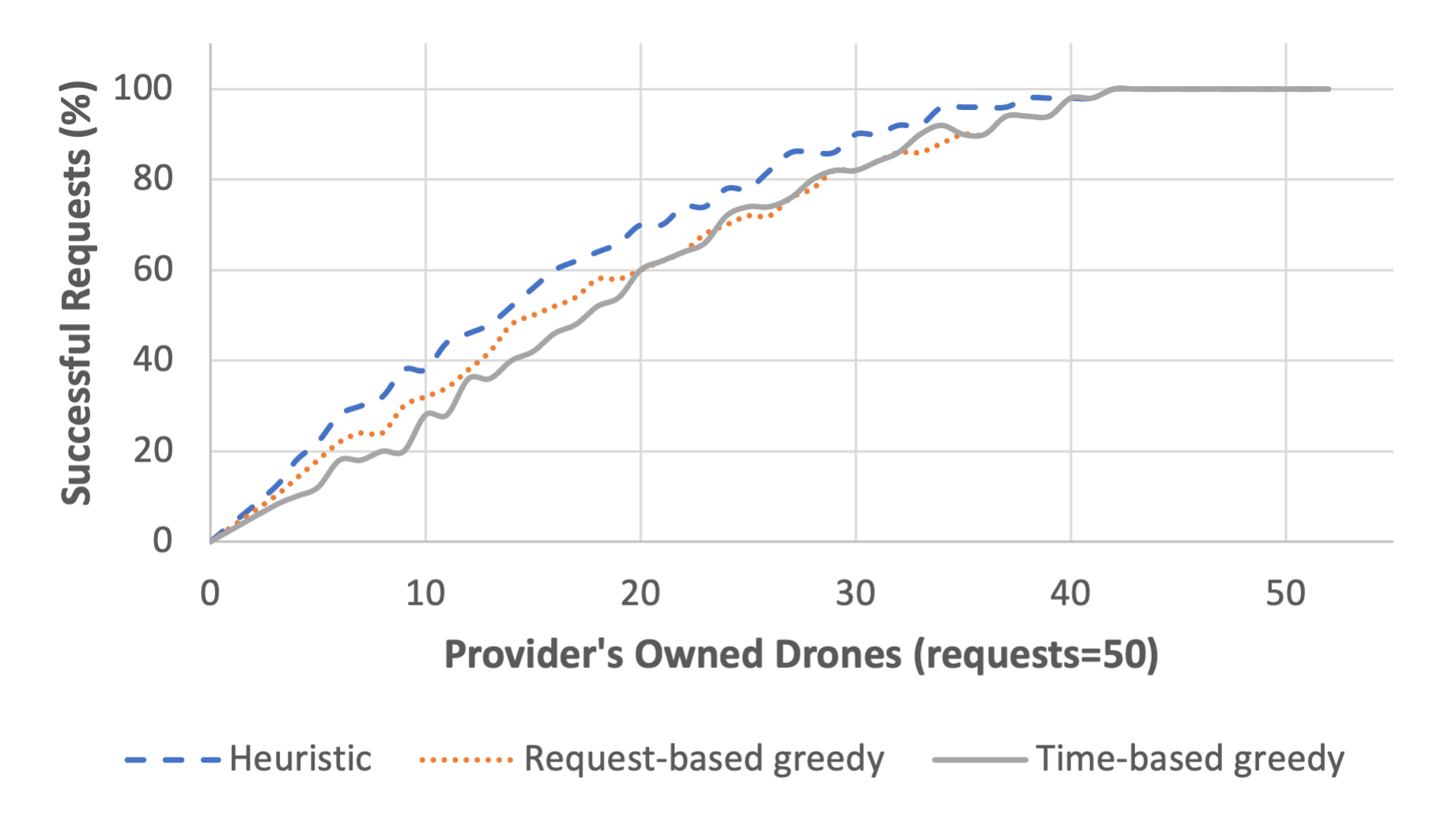}}
\caption{Successful requests varying the number of owned drones}
\label{successful-drones}
\end{figure}

In the second experiment, we study the performance of the proposed algorithms on the percentage of fulfilled requests. A fulfilled request is a request that is successfully allocated to a swarm of drones to be served. Fig. \ref{successful-requests} shows the percentage of successfully fulfilled requests as the number of requests increase in a day. We assume that the provider owns a set of 30 drones only. As shown in the diagram, when the number of requests are low the provider's drones are able to serve all the requests. As the requests received per day increase, the percentage of successful requests decreases. This is because the finite drones limit the number of requests that could be served. The heuristic-based approach distinctly outperforms the greedy approaches and is capable of serving more requests a day. This is reflected by the profit the provider gains (Fig. \ref{profit-requests}) and will lead to more customer's satisfaction as more requests are served. Such graph, allows the provider to determine the number of requests they should receive a day if they own a finite number of drones and are incapable of enlarging the fleet. Fig. \ref{successful-drones} measures the percentages of fulfilled requests with a fixed number of requests (50) and varied number of provider's owned drones. The graph shows how the heuristic-based method outperforms the other two methods. As the number of owned drones increase more requests are getting fulfilled. Such graph, helps the provider to determine the optimal number of drones they should own if they receive a daily $r$ number of requests.

\begin{figure}[htbp]
\centerline{\includegraphics[width=3.5in]{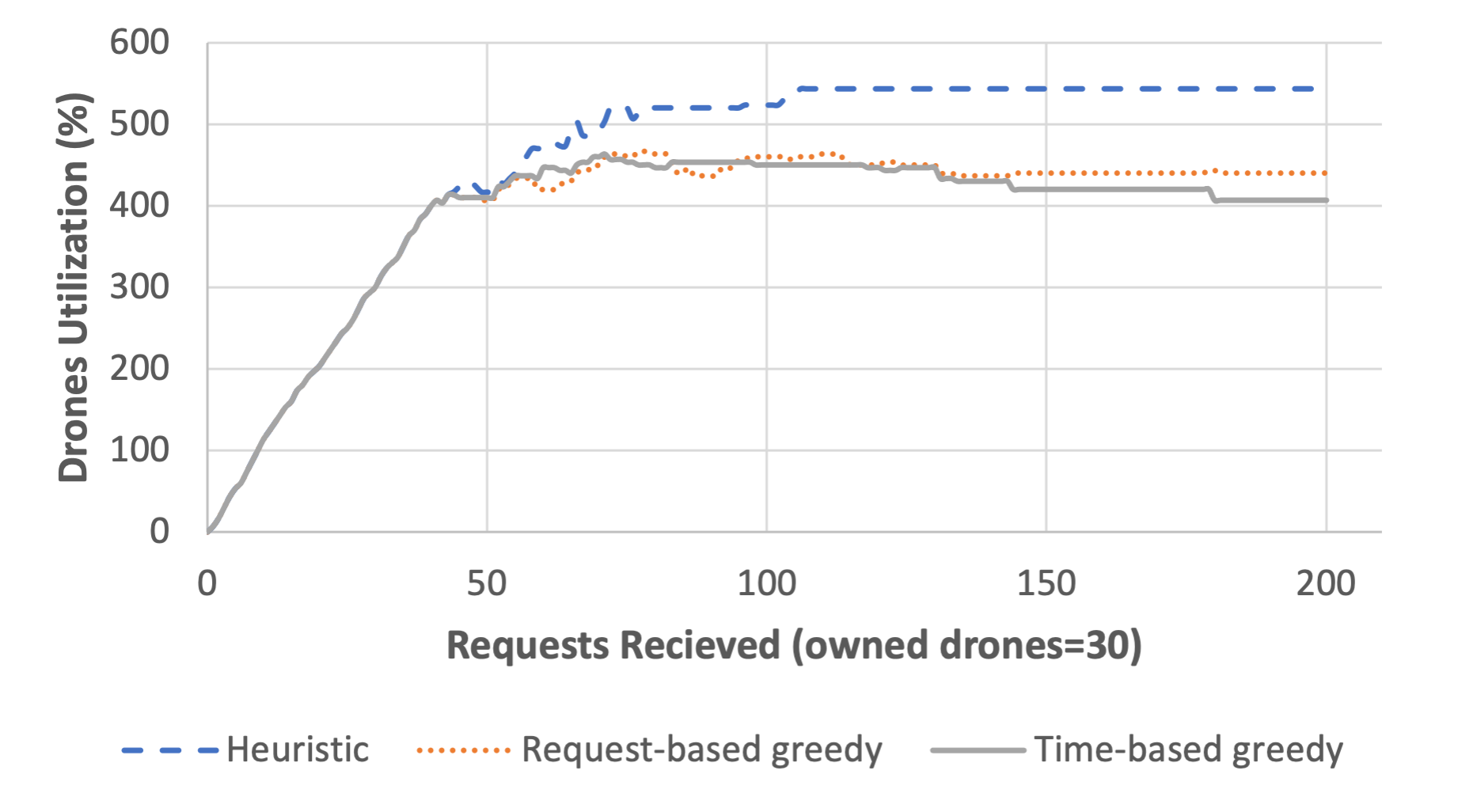}}
\caption{Drones Utilization varying the number of received requests}
\label{drones-requests}
\end{figure}

\begin{figure}[htbp]
\centerline{\includegraphics[width=3.5in]{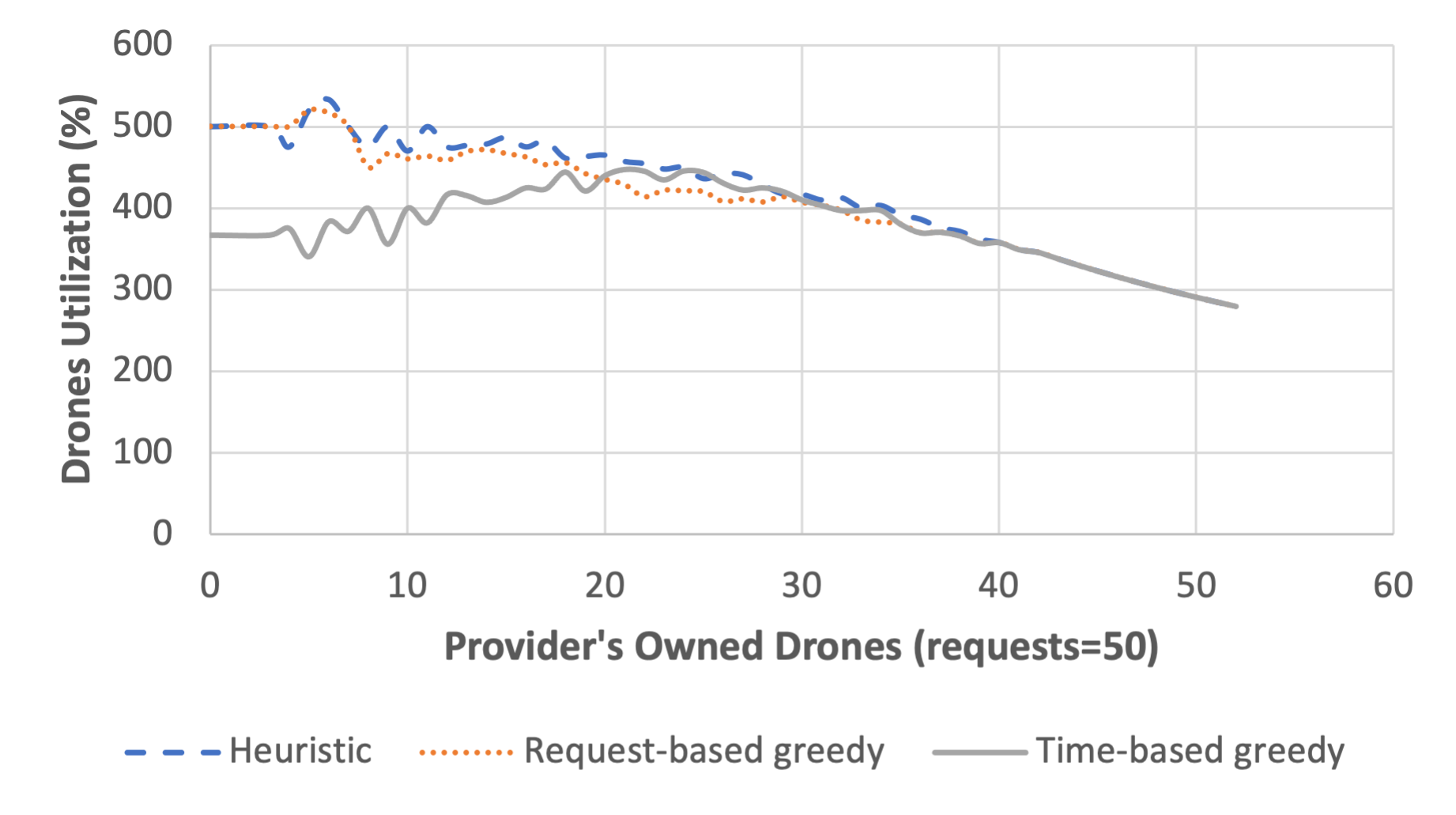}}
\caption{Drones Utilization varying the number of owned drones}
\label{drones-drones}
\end{figure}

In the last experiment, we measure the effectiveness of the methods in terms of drones utilization. A provider, as described earlier, owns a finite set of drones. An optimal method is a method that re-uses these drones as much as possible to serve multiple requests. Fig.\ref{drones-requests} shows the percentage of utilized drones, i.e. the number of times the drones are re-allocated, as the number of requests received increase. We assume that the provider owns 30 drones. As shown in the figure, as the number of received requests increase, the percentage of drone's utilization increases. The percentage stabilizes when all the drones are fully occupied at all time slots. As shown in the graph, the heuristic method outperforms the other methods in terms of drones utilization. The drones at requests 110-200 are used more than five times in a day with seven time windows and window overlapping requests. Fig. \ref{drones-drones} shows the percentages of drones utilized as the number of owned drones increases with a fixed number of received requests a day (50). As the number of owned drones increases the drones utilization decreases as they will be re-allocated less. The heuristic-based method, as shown, outperforms the other two methods in terms of drones utilization. At number of drones 5 the heuristic-based method and the request-based greedy algorithms utilize the same number of drones. However, if we look at Fig. \ref{successful-drones}, the percentage of fulfilled requests with 5 drones is less with the request-based greedy method. This means the request-based greedy method is using more drones to serve less number of requests. Using the drones utilization graphs along with the graphs of successful requests, the provider may determine the optimal number of drones they should own to serve maximum number of requests while utilizing the drones as much as they can without increasing the cost on themselves.

\section{conclusion}
We proposed a provider-centric allocation of drone swarm services known as, Swarm-based Drone-as-a-Service (SDaaS). A congestion-aware SDaaS composition algorithm is proposed to compute the maximum round trip times a swarm may take to serve a request taking the constraints at intermediate nodes (limited recharging pads and congestion) in consideration. Three swarms-requests allocation  methods were proposed with the goal of increasing the provider's profit namely; request-based greedy, time-based greedy, and heuristic-based allocations. The efficiency of the proposed approaches was evaluated in terms of profit maximization, execution times, requests fulfilment, and drones utilization. The efficiency of the proposed approaches where compared to a brute force baseline approach and has shown their feasibility on scaling compared to the brute force. Experimental results show the outperformance of the heuristic-based approach in comparison with the other two algorithms and its near optimal results. The limitation of the heuristic-based method is also described and compared to the brute force method. In the future work, we will consider heterogeneous swarms allocation to serve multiple requests and extend the work to deal with SDaaS failures.  




\bibliographystyle{ieeetr}
\bibliography{scholar}

\begin{thebibliography}{10}

\bibitem{shahzaad2021resilient}
B.~Shahzaad, A.~Bouguettaya, S.~Mistry, and A.~G. Neiat, ``Resilient
  composition of drone services for delivery,'' {\em FGCS}, 2021.

\bibitem{mogili2018review}
U.~R. Mogili and B.~Deepak, ``Review on application of drone systems in
  precision agriculture,'' {\em Procedia computer science}, vol.~133, 2018.

\bibitem{madawalagama2016low}
S.~Madawalagama, N.~Munasinghe, S.~Dampegama, and L.~Samarakoon, ``Low cost
  aerial mapping with consumer grade drones,'' in {\em 37th Asian Conference on
  Remote Sensing}, 2016.

\bibitem{shahzaad2020game}
B.~Shahzaad, A.~Bouguettaya, and S.~Mistry, ``A game-theoretic
  drone-as-a-service composition for delivery,'' in {\em 2020 ICWS}, IEEE.

\bibitem{forum_2020}
W.~E. Forum, ``We're about to see the golden age of drone delivery – here's
  why,'' Jun 2020.

\bibitem{euchi2020drones}
J.~Euchi, ``Do drones have a realistic place in a pandemic fight for delivering
  medical supplies in healthcare systems problems,'' {\em Chin. J. Aeronaut},
  2020.

\bibitem{alkouz2020swarm}
B.~Alkouz, A.~Bouguettaya, and S.~Mistry, ``Swarm-based drone-as-a-service
  (sdaas) for delivery,'' in {\em 2020 IEEE International Conference on Web
  Services (ICWS)}, pp.~441--448, IEEE, 2020.

\bibitem{cardona2019robot}
G.~A. Cardona and J.~M. Calderon, ``Robot swarm navigation and victim detection
  using rendezvous consensus in search and rescue operations,'' {\em Applied
  Sciences}, vol.~9, no.~8, p.~1702, 2019.

\bibitem{waibel2017drone}
M.~Waibel, B.~Keays, and F.~Augugliaro, ``Drone shows: Creative potential and
  best practices,'' tech. rep., ETH Zurich, 2017.

\bibitem{cao2018airborne}
X.~Cao, P.~Yang, M.~Alzenad, X.~Xi, D.~Wu, and H.~Yanikomeroglu, ``Airborne
  communication networks: A survey,'' {\em IEEE Journal on Selected Areas in
  Communications}, vol.~36, no.~9, pp.~1907--1926, 2018.

\bibitem{alkouz2020formation}
B.~Alkouz and A.~Bouguettaya, ``Formation-based selection of drone swarm
  services,'' {\em EAI Mobiquitous Conference}, 2020.

\bibitem{thornton2018swarming}
S.~Thornton and G.~E. Gallasch, ``Swarming logistics for tactical last-mile
  delivery,'' in {\em International Conference on Science and Innovation for
  Land Power}, vol.~2018, 2018.

\bibitem{van2008particle}
J.~van Ast, R.~Babu{\v{s}}ka, and B.~De~Schutter, ``Particle swarms in
  optimization and control,'' {\em IFAC Proceedings Volumes}, 2008.

\bibitem{chen2020toward}
W.~Chen, J.~Liu, H.~Guo, and N.~Kato, ``Toward robust and intelligent drone
  swarm: Challenges and future directions,'' {\em IEEE Network}, 2020.

\bibitem{shrit2017new}
O.~Shrit, S.~Martin, K.~Alagha, and G.~Pujolle, ``A new approach to realize
  drone swarm using ad-hoc network,'' in {\em 16th Annual Mediterranean Ad Hoc
  Networking Workshop}, pp.~1--5, IEEE, 2017.

\bibitem{hildmann2019nature}
H.~Hildmann, E.~Kovacs, F.~Saffre, and A.~Isakovic, ``Nature-inspired drone
  swarming for real-time aerial data-collection under dynamic operational
  constraints,'' {\em Drones}, vol.~3, no.~3, p.~71, 2019.

\bibitem{san2016delivery}
K.~T. San, E.~Y. Lee, and Y.~S. Chang, ``The delivery assignment solution for
  swarms of uavs dealing with multi-dimensional chromosome representation of
  genetic algorithm,'' in {\em IEEE 7th Annual UEMCON}, pp.~1--7, IEEE, 2016.

\bibitem{kuru2019analysis}
K.~Kuru, D.~Ansell, W.~Khan, and H.~Yetgin, ``Analysis and optimization of
  unmanned aerial vehicle swarms in logistics: An intelligent delivery
  platform,'' {\em IEEE Access}, vol.~7, pp.~15804--15831, 2019.

\bibitem{khamis2015multi}
A.~Khamis, A.~Hussein, and A.~Elmogy, ``Multi-robot task allocation: A review
  of the state-of-the-art,'' {\em Cooperative Robots and Sensor Networks 2015},
  pp.~31--51, 2015.

\bibitem{gigliotta2018equal}
O.~Gigliotta, ``Equal but different: Task allocation in homogeneous
  communicating robots,'' {\em Neurocomputing}, vol.~272, pp.~3--9, 2018.

\bibitem{dorigo2013swarmanoid}
M.~Dorigo, D.~Floreano, L.~M. Gambardella, F.~Mondada, S.~Nolfi, T.~Baaboura,
  M.~Birattari, M.~Bonani, M.~Brambilla, A.~Brutschy, {\em et~al.},
  ``Swarmanoid: a novel concept for the study of heterogeneous robotic
  swarms,'' {\em IEEE Robotics \& Automation Magazine}, 2013.

\bibitem{bouguettaya2017service}
A.~Bouguettaya, M.~Singh, M.~Huhns, Q.~Z. Sheng, H.~Dong, Q.~Yu, A.~G. Neiat,
  S.~Mistry, B.~Benatallah, B.~Medjahed, {\em et~al.}, ``A service computing
  manifesto: the next 10 years,'' {\em Communications of the ACM}, vol.~60,
  no.~4, pp.~64--72, 2017.

\bibitem{akram2017security}
R.~N. Akram, K.~Markantonakis, K.~Mayes, O.~Habachi, D.~Sauveron, A.~Steyven,
  and S.~Chaumette, ``Security, privacy and safety evaluation of dynamic and
  static fleets of drones,'' in {\em IEEE/AIAA 36th DASC}, 2017.

\bibitem{dijkstra1959note}
E.~W. Dijkstra {\em et~al.}, ``A note on two problems in connexion with
  graphs,'' {\em Numerische mathematik}, vol.~1, no.~1, pp.~269--271, 1959.

\bibitem{carlier1989algorithm}
J.~Carlier and {\'E}.~Pinson, ``An algorithm for solving the job-shop
  problem,'' {\em Management science}, vol.~35, no.~2, pp.~164--176, 1989.

\bibitem{salkin1975knapsack}
H.~M. Salkin and C.~A. De~Kluyver, ``The knapsack problem: a survey,'' {\em
  Naval Research Logistics Quarterly}, vol.~22, no.~1, pp.~127--144, 1975.

\bibitem{karduni2016protocol}
A.~Karduni, A.~Kermanshah, and S.~Derrible, ``A protocol to convert spatial
  polyline data to net. formats and apps to world urban road nets.,'' {\em
  Scientific data}, vol.~3, p.~160046, 2016.

\end{thebibliography}

\end{document}